\documentstyle[preprint,aps,epsfig]{revtex}
\begin{document}
\draft
\pagestyle{plain}
\newcommand{\D}{\displaystyle}
\title{\bf Effects of Fermi Surface Anisotropy on 
Unconventional Superconductivity in
${\bf U\!Pt_{3}}$}
\author{G. Hara\'n} 
\address{Institute of Physics, Politechnika Wroc{\l}awska, 
Wyb.Wyspia\'nskiego 27, 50-370 Wroc{\l}aw, Poland}
\author{P. J. Hirschfeld} 
\address{Department of Physics, University of Florida, 
Gainesville, FL 32611, USA}
\author{and M. Sigrist\cite{AA}}
\address{Theoretische Physik, ETH, 8093 Z\"urich, Switzerland }
\date{November 6, 1997}
\maketitle

\begin{abstract}
We discuss the weak-coupling BCS theory of the heavy fermion superconductor 
${\rm UPt}_3$, accounting for that system's anisotropic, multisheeted 
Fermi surface by expanding the order parameter and pair potential in 
terms of appropriate basis functions of the irreducible representations of 
the $D_{6h}$ crystal point group. Within a phenomenological model for 
the electronic structure of ${\rm UPt}_3$ chosen to capture the qualitative 
features of local density functional calculations and de Haas-van Alphen  
measurements, we show how Fermi surface anisotropy can favor pairing 
in certain symmetry classes, and influence the phase diagram of unconventional  
superconductors. 
We also calculate the Ginzburg-Landau coefficients, 
focussing on those coefficients relevant to current theories of the 
${\rm UPt}_3$ phase diagram.
\end{abstract} 

\section{Introduction}
Several different scenarios have been proposed \cite{1,2,3,4,5} to explain the
unusual phase diagram of the heavy fermion superconductor 
${\rm UPt}_3$ in an external magnetic field \cite{6,7} and under pressure. \cite{8,9} 
This phase diagram consists of several superconducting
phases of presumably
different symmetry. Almost all theories of the phase diagram
involve unconventional superconductivity, i.e. superconducting
order parameters of lower symmetry than the conventional one
which only breaks the U(1) gauge symmetry. These scenarios have 
almost exclusively been discussed on the basis of generalized
Ginzburg-Landau (GL) theories of superconductivity, where the free
energy functional is derived purely by symmetry arguments. \cite{10,11,12,13} 
It is the nature of this approach that such theories
require a considerable number of
adjustable phenomenological parameters. In addition the
choice of the order parameter, which can be
classified according to the irreducible representations
of the normal state symmetry group of the system, is open. Each
representation corresponds to a Cooper pairing channel with a
particular pairing energy, i.e. transition temperature.

At present, no satisfactory microscopic theory exists which could
assist in picking out the relevant representation and fixing
the phenomenological parameters of the corresponding GL theory.
Of course, it is possible to work based on semimicroscopic
formulation of a generalized weak-coupling BCS theory. This
type of approach is often used in the study of properties
related to the energy spectrum of the quasiparticle excitations
for unconventional superconducting states with a specific 
topology of nodal structure of the quasiparticle gap, and
considerable insight for various low-temperature properties has
been gained from these treatments. \cite{13} However, they are insufficient
to answer the open questions we have in determining a GL theory. 
One difficulty arises from the fact that the specific form of the attractive
pairing potential is unknown. Various additional corrections to
the simplified weak-coupling treatment might be 
important and need to be properly taken into account. One aspect is 
the shape of the Fermi surface, (FS)  
which has multiple sheets, some of them very anisotropic. \cite{14,15,16} 
In conventional weak-coupling calculations the FS has usually 
a spherical symmetry.
It would be helpful to connect the semimicroscopic and the GL approach
by including effects of the FS anisotropy into the weak-coupling
approach. Such effects are an important part in determining the
symmetry of the relevant
Cooper pairing channel as well as the values of
the parameters in the GL theory, and have so far not been
considered in the literature. 

Let us now briefly review the situation in the theory of $ {\rm UPt}_3 $. 
The scenarios for the $ {\rm UPt}_3 $-phase diagram can be divided into
three classes. (1) The relevant order parameter belongs to a two-dimensional (2D)
representation of the crystal point group symmetry $ D_{6h} $ (hexagonal) of   
${\rm UPt}_3$, $ E_{1(g,u)} $ or $ E_{2(g,u)} $. \cite{1,2,17,18} In this
scenario a symmetry  lowering field is required to lift the two-fold
degeneracy of the 2D order parameter in order to describe the experimentally
observed double transition. Such a field is given, for example, by the
staggered moment of the weak antiferromagnetic order which appears below
$ T_N \approx 5K $ in $ {\rm UPt}_3 $. \cite{22} 
To this class of theories belongs in spirit an idea due to Zhitomirskii and
Luk'yanchuk, \cite{5} which states that the system is actually 
close to an isotropic
one.  Hexagonal anisotropy is then taken to split the representations of the
rotation group, leaving nearly degenerate 1D or 2D representations.
(2) The double transition can also 
occur for a triplet order parameter described by the one-dimensional (1D) representations 
of $D_{6h}$ group, when the spin-orbit coupling interaction is negligible. In this 
model the degeneracy due to the spin direction is lifted by the staggered magnetic 
moment. \cite{4,19} (3) It is also plausible, that two irreducible representations  
of $ D_{6h} $ might have their transition temperatures very close to
each other (accidental near degeneracy) such that they would generate the
feature of the double transtion. \cite{3,5} The staggered moment does not play any
role in this type of scenario. 

As a test case for these scenarios usually the tetracritical point in
the $ H $-$ T $-phase diagram of $ {\rm UPt}_3 $ at $ (H,T) \approx
(500 Gauss, 0.4 K) $ is analyzed. This point is most clearly observed 
for fields parallel to the basal plane, and is rather insensitive
to the relative orientation with respect to the basal plane crystalline
axes. The scenarios (1) have, in general, more difficulty than the 
scenarios (3) to reproduce
this detail because of the anisotropy introduced by the staggered moment.   
One argument to overcome this flaw is the assumption that the staggered
moment always follows the orientation of the applied magnetic field.
A further problem for the type 1) theories is that  
the observed tetracritical point
for fields along the c-axis is not well reproduced. 
Sauls suggests that the problem could also be solved if
certain parameters in the GL theory are very small. \cite{1} Based on this idea
he gave an argument in favor of the irreducible representation 
$E_{2u}$ (f-wave) which is a triplet pairing state.  Joynt\cite{ParkJoynt} argues
similarly that in an $E_{1g}$ 
representation an  apparent tetracritical point for 
${\bf H}\parallel \hat c$ is obtained if the system is nearly particle-hole
symmetric.  The data for this configuration are not conclusive.\cite{6}
The scenarios (2) of the order parameter given by 1D 
representations and (3) which neglects the coupling between
the staggered moment and the superconducting order parameter predict
a tetracritical point in all directions.

Strong evidence in favor of theories (types 1) \& 2)) which rely on the
antiferromagnetic order as a symmetry breaking field is the observed simultaneous
disappearance of both the high temperature, low field ``A phase" and the staggered
moment with the application of hydrostatic pressure.\cite{8,9}  The disappearance of
the two transitions could be taken as evidence for the Zhitomirskii-Luk'yanchuk
scenario, suggesting a nearness to ``isotropy".  However the simultaneous
vanishing of the staggered  magnetization seems more natural in the context of the 
2D theories based on a symmetry breaking field.  
More recently, it was found that at the pressure value where the transitions
merge, application of basal plane stress ``resplit" the transition, lending
further support to these scenarios.\cite{Rosenbaum}  On the other hand, in these
theories the staggered magnetization generally must be assumed to rotate
with the applied field in order to understand both the isotropy of the
phase diagram for ${\bf H} \perp \hat c$ and the weak sixfold variation
of the upper critical field $H_{c2}$ with angle in the plane.\cite{Sauls}
The recent experiment of Lussier et al. which found no rotation of the staggered
moment with field appears to contradict this hypothesis, however.\cite{Lussier}

Further constraints on the theory are provided by recent Knight shift measurements,
which show no temperature dependence for any direction of the field below 
$T_c$. \cite{Tou} This striking result would be consistent with an equal spin pairing
state such as in $^3He-A$, with Cooper pair spins free to rotate with the
external field.  Such a model (type 3), in which the order parameter is a
spin triplet state with 3 spin components split by the staggered moment,
has indeed been proposed by Machida and co-workers.\cite{4,newMachida}  This
theory has the disadvantage of generically exhibiting three transitions in
zero field rather than two, due to the three independent order parameter 
components.   It also cannot explain the apparent Pauli limiting of the upper
critical field for ${\bf H}\parallel \hat c$.\cite{ChoiSauls}

Thus, the existing experimental data have not been particularly kind to
any of the current theoretical scenarios.  It is not our aim to identify the
correct theory here.  Rather, we focus on an aspect which has been generally
neglected in oversimplified models, namely which types of order parameter
might be favored by the unusual, highly anisotropic Fermi surface of $UPt_3$.
Within an extension of the semimicroscopic 
weak-coupling theory, we hope to be able to suggest 
which scenario is most likely, although no concrete proof can be given.
From our analysis we expect to answer or shed light on some of the questions
of interest. Are two nearly degenerate order parameters very likely
to appear in $ {\rm UPt}_3 $? Does the scenario (1) given by Sauls
satisfy the conditions that GL theory would produce the isotropic
tetracritical point?  Is the system close to ``isotropy" in the sense
of Zhitomirskii and Luk'yanchuk?  Which representations are favored by
hexagonal anisotropy?  And, how is the pairing potential distributed among
the various sheets of the multisheeted Fermi surface?

Our discussion is considerably simplified by the assumption
that a weak-coupling type of theory can be applied in case
of $ {\rm UPt}_3 $. Several properties of $ {\rm UPt}_3 $ support
the validity of such an approach. First, $ {\rm UPt}_3 $ (in 
contrast to $ {\rm UBe}_{13} $) is in a rather well-developed
Fermi liquid (FL) regime at the onset of superconductivity. For 
example, the linear specific heat and the $ T^2 $-dependence
of the resistivity are suggestive for the existence of well-defined
(Landau) quasiparticles. \cite{14,23} Their effective mass is extremely large
such that the degeneracy temperature $ T_F $ is of the order
of $ 10 \sim 20 K $. The question arises whether strong coupling
effects are important in such a situation. They are measured by the
ratio $ T_c / T_F $ which is here of the order of 1/20 . Thus we
neglect these corrections, although they are considerably stronger
than in conventional superconductors with $ T_c / T_F \sim 10^{-3} $.
Indirect justification for this assumption is provided by the size
of the specific heat discontinuity at the upper transition point,
$\Delta C/C_N\sim 0.8$, \cite{20} which has to be compared with the BCS 
weak-coupling result, $ \Delta C/C_N\approx 1.43 $. The FL renormalizations
are of course quite large, as evidenced by the large effective mass.
However, we will focus our attention to the
properties of the superconductor close to the transition temperature,
where these corrections are irrelevant. \cite{21} Furthermore we simplify
our discussion by taking the pairing potential as the only interaction
among the quasiparticles into account. Thus the only modification
to the usual weak-coupling theories is inclusion of the Fermi
surface anisotropy. This implies also that pair correlation can be
confined to a narrow energy shell about the FS of a
width $ \omega_c $, the cutoff energy associated with the 
electron-electron interaction (analogous to the Debye frequency
$ \omega_D $ in conventional superconductors).

\section{Fermi surface anisotropy and the choice of the
order parameter}

\subsection{The linearized gap equation}
The BCS description of the superconductor
is based on the following typical Hamiltonian

\begin{equation}
\label{e1}
{\cal H}= \sum_{{\bf k},s} \varepsilon_{\bf k} c^{\dag}_{{\bf k}s} 
c_{{\bf k}s}
+ \frac{1}{2}\sum_{{\bf k},{\bf k}'} \sum_{s_1,..,s_4}
V_{s_1 s_2 s_3 s_4}({\bf k},{\bf k}')
c^{\dag}_{-{\bf k}s_1} c^{\dag}_{{\bf k}s_2}
c_{{\bf k}'s_3} c_{-{\bf k}'s_4}
\end{equation}

\noindent
where $ c^{\dag}_{{\bf k}s} $ ($ c_{{\bf k}s} $) denotes the
creation (annihilation) operator of a quasiparticle with
momentum $ {\bf k} $ and spin $ s $. The quasiparticle band
$ \varepsilon_{\bf k} $ is measured relative to the Fermi
energy $ \varepsilon_F $ and $ V_{s_1 s_2 s_3 s_4}({\bf k},{\bf k}') $
is the pairing potential which is finite only for momenta
$ {\bf k} $ and $ {\bf k}' $ for which $ |\varepsilon_{{\bf k}}|
< \omega_c $, $ |\varepsilon_{{\bf k'}}|< \omega_c $, where only scattering processes
between particles of opposite momentum are included. 
The superconducting state is
expressed by the mean field

\begin{equation}
\label{e2}
\Delta_{{\bf k}s_1 s_2} = - \sum_{{\bf k}'} \sum_{s_3,s_4}
V_{s_1 s_2 s_3 s_4} ({\bf k},{\bf k}') \langle 
c_{{\bf k}'s_3} c_{-{\bf k}'s_4} \rangle
\end{equation}

\noindent
where $ \langle .... \rangle $ is the thermal expectation
value. The mean field $ \Delta_{{\bf k}ss'} $ describes the
gap of the quasiparticle spectrum
$ E_{{\bf k}} = [\varepsilon^2_{{\bf k}}
+ tr( \hat{\Delta}^{\dag}_{{\bf k}}
\hat{\Delta}_{{\bf k}})/2]^{1/2} $. The gap function 
$ \hat{\Delta}_{{\bf k}} $
is a $ 2 \times 2 $-matrix in spin space. By inserting the expectation
value for a given temperature
$T$ in Eq. (2) we obtain the self-consistency equation for
the gap function,

\begin{equation}
\label{e3}
\Delta_{{\bf k} s_1 s_2} = - \sum_{{\bf k}'} \sum_{s_3 , s_4}
V_{s_1 s_2 s_3 s_4}({\bf k}, {\bf k}') 
\frac{\Delta_{{\bf k}' s_3 s_4}}{2 E_{{\bf k}}}
{\rm tanh}\left( \frac{E_{{\bf k}}}{2 k_B T} \right)
\end{equation}

\noindent
This equation can be used to
determine $ T_c $ by considering the limit of the order parameter amplitude 
$ \Delta \to 0 $ which yields the linearized gap equation 

\begin{equation}
\label{e3a}
\nu\Delta_{{\bf k} s_1 s_2} = - \sum_{{\bf k}'} \sum_{s_3 ,s_4}
V_{s_1 s_2 s_3 s_4} ({\bf k}, {\bf k}') 
\Delta_{{\bf k}' s_3 s_4} \delta(\varepsilon_{{\bf k}})
\end{equation}

\noindent
where $ 1/\nu = {\rm ln}(1.14 \omega_c /k_B T_c) $ is obtained
by integrating $ {\rm tanh}(\varepsilon/2 k_B T_c)/2 \varepsilon $
over $ \varepsilon $ in the energy shell, $ - \omega_c < \varepsilon
< \omega_c $. This equation is an eigenvalue equation where the
largest among the eigenvalues $ \nu $ determines the superconducting
transition temperature $ T_c $ and the corresponding symmetry
of gap function.

The Cooper pairing states can be classified into even (spin singlet) 
and odd (spin triplet) parity states which can be parametrized
by a scalar function

\begin{equation}
\label{e4}
\hat{\Delta}_{\bf k}= \Delta\psi ({\bf k}) i\hat{\sigma}_y
\end{equation}

\noindent
for even parity and a vector function

\begin{equation}
\label{e5}
\hat{\Delta}_{{\bf k}} = \Delta\sum_{\mu=x,y,z} d_{\mu}
({\bf k}) i\hat{\sigma}_{\mu}\hat{\sigma}_y
\end{equation}

\noindent
for odd parity. If we assume that in the scattering described by
$ V_{s_1 s_2 s_3 s_4} ({\bf k}, {\bf k}') $ spins are not affected
then the pairing potential has the following form

\begin{equation}
\label{e6}
V_{s_1 s_2 s_3 s_4} ({\bf k}, {\bf k}') = \tilde{V} ({\bf k}, {\bf k}')
(\delta_{s_1 s_3} \delta_{s_2 s_4} \pm \delta_{s_1 s_4} \delta_{s_2 s_3})
\end{equation}

\noindent
where the $ + $ ($ - $)-sign occurs for 
$ V_{s_1 s_2 s_3 s_4} ({\bf k}, {\bf k}') $ supporting 
odd (even) parity pairing. In this case the linearized gap
equation can be expressed in the new parametrization as

\begin{equation}
\label{e7}
\nu\psi({\bf k}) = - \sum_{{\bf k}'} \tilde{V}({\bf k}, {\bf k}')
\psi({\bf k}') \delta(\varepsilon_{{\bf k}'})
\end{equation}

\noindent
and

\begin{equation}
\label{e8}
\nu d_{\mu} ({\bf k}) = - \sum_{{\bf k}'} \tilde{V}({\bf k}, {\bf k}')
d_{\mu} ({\bf k}') \delta(\varepsilon_{{\bf k}'})
\end{equation}

\noindent
Note that in the case of odd parity pairing the different components
of $ d_{\mu} ({\bf k}) $ do not mix in the linearized gap equation.

\subsection{Order parameter}
In our weak-coupling approach we ignored quasiparticles further
away from the FS so that in the linearized
gap equation given in Eq. (4) the momentum sum is restricted to the 
Fermi surface. In this case we express the gap function in terms of 
the linear combinations of the spherical harmonics defined on an appropriate  
Fermi surface. 
It is however necessary to choose this set of polynomials 
$ \{ \phi_l ({\bf k}) \} $ so that they are 
orthonormal to each other in the sense that the
scalar product is determined as a FS average

\begin{equation}
\label{e9}
\sum_{{\bf k}} \phi_l ({\bf k} )
\phi_{l'} ({\bf k}) 
\delta(\varepsilon_{{\bf k}}) = \delta_{ll'}
\end{equation}

\noindent
where $ l $ is a label for the polynomial.
This property is most easily achieved by classifying them 
with respect to the irreducible representations
of the crystal field point group, $ D_{6h} $ in our case.
Then only the polynomials within each irreducible representation have to
be orthogonalized. 
 
The gap function for the singlet state can now be written as

\begin{equation} 
\label{e10}
\psi({\bf k}) = \sum^n_{l=1} \eta_l \phi_{l} ({\bf k})
\end{equation}

\noindent
where $ \eta_l $ are the complex expansion coefficients.
For the triplet state we have to express the vector $ {\bf d}({\bf k})$
which requires an additional set of polynomials $ \{ \varphi^{\mu}_l
({\bf k} ) \} $ which leads to

\begin{equation}
\label{e11}
d_{\mu} ({\bf k}) = \sum^n_{l=1} \eta_{\mu l} \varphi^{\mu}_l 
({\bf k}) 
\end{equation}

\noindent
Orthogonality in this case is defined by

\begin{equation}
\label{e12}
\sum_{{\bf k}} \sum_{\mu=x,y,z} \varphi^{\mu}_l ({\bf k})
\varphi^{\mu}_{l'}({\bf k}) \delta(\varepsilon_{{\bf k}})
= \delta_{ll'}
\end{equation}

\noindent
In Tabs. I-III we catalogue these polynomials up to the order 3. We can now
use them to express the linearized gap equation in matrix form as we
show here for the case of even parity

\begin{equation}
\label{e13}
\nu\eta_l = - \sum_{l'} \langle l | \tilde{V} | l'\rangle \eta_{l'}
\end{equation}

\noindent
where

\begin{equation}
\label{e14}
\langle l | \tilde{V} | l' \rangle = \sum_{{\bf k},{\bf k}'} 
\delta(\varepsilon_{{\bf k}})\delta(\varepsilon_{{\bf k}'}) 
\phi_{l} ({\bf k})
\tilde{V}({\bf k}, {\bf k}') \phi_{l'} ({\bf k}')
\end{equation}

\noindent
Clearly, this eigenvalue equation decays into block form
where each block belongs to one irreducible representation
of $ D_{6h} $. Thus we may consider the problem for each
representation separately. There are representations in
Tabs. I and III  which have more than one polynomial. It is necessary
to include all of them and ensure that they are orthonormalized.
Note that the orthonormalization depends on the properties of the FS.

Sometimes it is advantageous to express the gap function in 
terms of so-called FS harmonics, as introduced by Butler and Allen,\cite{25,26}
and 
discussed in the context of the $UPt_3$ pairing problem
in Ref. \cite{33}.   The Fermi surface harmonics are homogeneous polynomials 
of the Fermi velocity, $ {\bf v}_{{\bf k}} = \nabla_{{\bf k}} 
\varepsilon_{{\bf k}} $ and in the case of $D_{6h}$ point group they 
can be obtained from Tabs. I-III by replacing ${\bf k}$ ($k_i$) by 
${\bf v}_{{\bf k}}$ ($v_i$). These functions 
are to be orthonormalized in the same way as the polynomials  
constructed from spherical harmonics. We will use the FS harmonic representation 
of the order parameter in Section 3 of this paper. 

\subsection{Pairing interaction and electron band}
Now we have to specify the pairing interaction $ \tilde{V}({\bf k},
{\bf k}') $ and the electron band $ \varepsilon_{{\bf k}} $.
In a spherical symmetric system the pairing states would be
classified according to their angular momentum quantum numbers,
$ \ell $ and $ m $ (and radial quantum numbers among which
we will consider here only the lowest ones
in each angular momentum channel). States for a given $ \ell $
are degenerate, i.e. yield in the linearized gap equation above
the same transition temperature $ T_c $. It is natural then
to write the corresponding pair potential as

\begin{equation}
\label{e15}
V_{s_1 s_2 s_3 s_4}({\bf k}, {\bf k}') = \sum_{\ell} V_{\ell}
\sum^{+\ell}_{m=-\ell} Y_{\ell m}({\bf k}) 
 Y^*_{\ell m}({\bf k}') [\delta_{s_1 s_3} \delta_{s_2 s_4}
- (-1)^{\ell} \delta_{s_1 s_4} \delta_{s_2 s_3}]
\end{equation}

\noindent
where $ Y_{\ell m} $ are the spherical harmonics depending on the Fermi 
vectors ${\bf k}$. 
Note that the $ Y_{\ell m}({\bf k}) $ are dimensionless but are
different from the usual $ Y_{\ell m}({\hat k}) $ away from the isotropic
Fermi surface point.
In this notation $ V_{\ell}$ is negative for an attractive potential. 
In the following we will consider the FS anisotropy (crystal field) 
effects on the eigenvalues 
of Eq. (4). The information we can obtain is how 
the degenerate eigenstates evolve as the FS gradually
becomes anisotropic. We will keep
the interaction in this form where $ V_{\ell} $ is an unknown
parameter. 

Information from experimental data as well as band structure
calculation can help us to obtain 
an approximate electron band $ \varepsilon_{{\bf k}} $.
While the FS of $ {\rm UPt}_3 $ is highly complex and consists of 
at least five sheets, \cite{14,15,16} not all of them
are equally important. We will focus
here on two bands only which are supposed to be the ones 
with the largest and the third-largest  
contribution to the density of states, that is  
the $ \Gamma_3 $ and $ \Gamma_2 $ FS sheet respectively (notation from Ref. 16).
A simple but crude fit of the band structure data can be achieved by
the following form for $ \varepsilon_{{\bf k}} $

\begin{equation}
\label{e16}
\varepsilon_{{\bf k}} = a_1 k^2_z + a_2 (k^2_x + k^2_y)
+ a_3 (k^3_y - 3 k^2_x k_y)^2 - \varepsilon_F
\end{equation}

\noindent
where $ a_i $ and $ \varepsilon_F $
are fit parameters. The $ \Gamma_2 $ band
is fitted by an ellipsoidal $ \varepsilon_{{\bf k}} $ which is
rotationally symmetric about the c-axis, that is $a_3=0$ in Eq. (\ref{e16}). 
The hexagonal
anisotropy is larger for the $ \Gamma_3 $ band
which we approximate by a ``wrinkled'' ellipsoid with $ a_3 \neq 0 $.
Note, that the latter
band has additional pockets at Brioullin zone
boundary which are not reproduced by  
$ \varepsilon_{{\bf k}} $ from Eq. (\ref{e16}).
However, this form of $ \varepsilon_{{\bf k}} $ has the advantage 
that we can follow continuously the development of the anisotropic bands
starting from a spherical symmetric one 
($ a_1=a_2 \neq 0 $ and $ a_3=0 $).
The values of the parameters in $ \varepsilon_{{\bf k}} $
for the anisotropic band
are obtained using the ratios $ k_{F_{x}}/k_{F_{z}} $ and 
$ k_{F_{y}}/k_{F_{z}} $ from LDA calculations \cite{16} and by conserving 
the area of cross section
in the $ k_{y} $-$ k_{z} $ plane as observed in de Haas-van
Alphen measurements. \cite{14} Our best fit parameters are given in Tab. IV, 
and the FS cross sections by the symmetry planes are shown in Fig. 1. 
This fit leaves still the density of states integrated over the FS 
as an undetermined parameter. In the following we will express
various quantities in dimensionless units refering to
the integrated density of states.

\subsection{Solution of the linearized gap equation}
We address now the problem of the stability of the possible
pairing states in the given $ \Gamma_2 $- and $ \Gamma_3 $-band.
In reality, superconductivity is a combined effect of 
all FS sheets, because quasiparticle scattering among the
different sheets will couple their pairing amplitudes.
It would be not difficult to introduce such interband
coupling in the pairing interaction $ \tilde{V}({\bf k},{\bf k'}) $, but 
this would lead to additional unknown coupling parameters.
In the following discussion we will avoid this complication
and consider the effect of the two FS sheets separately
as if they were uncoupled.

We investigate the effect of the FS anisotropy on the
pairing states by continuously distorting a spherical
FS into the above defined FS sheets $ \Gamma_2 $ or
$ \Gamma_3 $ (Eq. (\ref{e16})). For the spherical FS all pairing states
of the same angular momentum $ \ell $ are degenerate. 
The degeneracy of these $ 2 \ell +1 $ states is lifted
when the FS deviates from the spherical shape, and they
can be classified by the irreducible representations of
the new, lower symmetry $ D_{6h} $. Restricting our
attention to only one value of $ \ell $ ($ \ell=1,2 $ and 3)
at a time we need only to take one coupling parameter, $ V_{\ell} $,
into account. This assumption may be too simplistic for realistic pair potentials, 
but we make it in the first analysis to isolate the effects of the Fermi surface 
anisotropy. We use an eigenvalue $\nu_{\ell}$ of Eq. (\ref{e3a}) on the spherical 
FS of the same integrated density of states, $N(0)$, as a reference value 
for $\nu$. Expressed in dimensionless units $ \nu' =\nu/\nu_{\ell} $ 
does not depend on $N(0)$, which is not determined within this approach.  
The distortion from a spherical FS that is the effect of a crystal field 
is included by the dimensionless linear parameter $t$ changing from 0 
to 1, with $t=0$ corresponding to the spherical Fermi surface of the 
radius adjusted to dHvA data \cite{14} and $t=1$ representing 
$\Gamma_{3}$ or $\Gamma_{2}$ 
Fermi sheet (Tab. IV). Therefore an increase in $t$ implies a higher 
hexagonal ($\Gamma_{3}$) or ellipsoidal ($\Gamma_{2}$) perturbation of 
the Fermi surface. The $t$-dependence of the energy band coefficients for 
the hexagonal FS is given by

\begin{equation}
\begin{array}{l}
a_1\left(t\right)/\left(\varepsilon_F c^2\right)=0.225-0.068t\\
a_2\left(t\right)/\left(\varepsilon_F c^2\right)=0.225+0.016t\\
a_3\left(t\right)/\left(\varepsilon_F c^6\right)=0.016t
\end{array}
\end{equation}

\noindent
and for the ellipsoidal Fermi sheet 

\begin{equation}
\begin{array}{l}
a_2\left(t\right)/\left(\varepsilon_F c^2\right)=0.666-0.200t\\
a_2\left(t\right)/\left(\varepsilon_F c^2\right)=0.666+0.285t\\
a_3\left(t\right)=0
\end{array}
\end{equation}

\noindent
where $c$ is the z-axis lattice constant (Tab. IV).  
The solutions of the linearized gap equation (Eq. (4)) for 
the p-wave, d-wave and f-wave pair potentials as the functions of the 
crystal field $t$ are shown in Fig. 2 for the case of pairing interaction 
on $\Gamma_3$ Fermi sheet only (hexagonal crystal field)  
and Fig. 3 for the case of pairing potential on $\Gamma_2$ Fermi sheet  
(ellipsoidal crystal field). We present the 
critical temperature $T_c$ for triplet states (Figs. 2a,2c,3a,3c) with ${\bf d} 
\parallel \hat{z}$ (Eq. (\ref{e5})) only as the states with ${\bf d}
\bot\hat{z}$ are degenerate with them according to Tabs. I and III. 
Particularly interesting is the result for the d-wave states with the 
pairing interaction on $\Gamma_{3}$ FS (Fig. 2b). We observe here a competition 
between $A_{1g}$ and $E_{1g}$ states as the hexagonal anisotropy of the 
Fermi sheet is increased. Finally, for our best fit $\Gamma_{3}$ FS ($t=1$) 
the transition temperature of $A_{1g}$ state ($T_c(A_{1g})$) is higher than that of 
$E_{1g}$ state ($T_c(E_{1g})$), but the system is very close to the degeneracy point  
($t\approx 0.85$) where $T_c(A_{1g})=T_c(E_{1g})$. Thus application of 
a hydrodynamic pressure which decreases the crystal field $t$ and restores 
"isotropy" may drive the system to a state with a single transition temperature 
$T_c=T_c(A_{1g})=T_c(E_{1g})$, consistent with the Zhitomirskii and Luk'yanchuk 
scenario. \cite{5} The role of the hexagonal crystal field is worth mentioning, 
since as we show in Fig. 3b the cylindrical crystal field lifts the spherical 
FS ($t=0$) degeneracy of d-wave states and splits the transition temperatures 
corresponding to different irreducible representations. Therefore the situation  
of near degenerate states $A_{1g}$ and $E_{1g}$ is not realized when the 
pairing mainly takes place on the ellipsoidal $\Gamma_{2}$ FS. Analysis 
of the f-wave states shows that the $E_{2u}$ irreducible representation, suggested 
to describe the superconducting state in Sauls' scenario, \cite{1,18,24} is 
the one with the highest transition temperature on $\Gamma_3$ FS but almost 
degenerate with $A_{1u}$ representation (Fig. 2c). On the other hand,
 the transition 
temperature of $A_{1u}$ state is significantly higher than that of $E_{2u}$
on the $\Gamma_2$ sheet of FS (Fig. 3c).

\section{Ginzburg-Landau Coefficients}
Direct calculation of the GL coefficients incorporating aspects of the
anisotropy of the FS is important for several reasons.  The question
of most current interest is the consistency of the Sauls $E_{2u}$ scenario
for the ${\rm UPt}_3$ phase diagram. \cite{1} This theory relies on the smallness 
of a particular
term in the GL gradient free energy, which may be shown to destroy the
isotropy of the tetracritical point in the 2D scenario.  In the particular
case of the $E_{2u}$ representation, this term may be shown to vanish in
cylindrical symmetry, and it is therefore hexagonal anisotropy alone
which is responsible for the disagreement between the predictions of the
$E_{2u}$ scenario for the phase diagram and experiment.  If the effect
of the anisotropy could be shown to be quite small, i.e. the system were
in this sense quite close to a cylindrically symmetric one, one would have
considerably more confidence in the theory.

Before discussing this point in some detail, we remark however that
other important physical information can be gleaned from a 
weak coupling calculation of the GL coefficients over the FS.
Experimentally measurable quantities 
at the transition depend directly on combinations of the GL coefficients, e.g.
for the specific heat jump, $\Delta C/C_N = \alpha^2/2\beta_1 T_c$.
If sufficient independent measurements can be made to pin down 
the coefficients individually, and if the correct representation is
known, comparison with weak-coupling calculations
including the correct Fermi surface could in principle allow one
to determine the size of strong-coupling corrections.  This might
in turn serve as a guide for theories of the ${\rm UPt}_3$ normal state.

The form of the GL free energy density has been given in several places, e.g.
Refs. 1,13.  For the 2D representations in hexagonal symmetry 
it takes the form
\begin{equation}
\label{e17}
\begin{array}{l}
f=\alpha |\bar{\eta}|^{2}+\beta_{1}|\bar{\eta}|^{4}+\beta_{2}
|\bar{\eta}\bar{\eta}|^{2}
+\gamma_{1}|\bar{\eta}|^{6}+\gamma_{2}|\bar{\eta}|^{2}
|\bar{\eta}\bar{\eta}|^{2}+\\
\\
\sum_{i,j=x,y}\{\kappa_{1}\left(\partial_{i}\eta_{j}\right)
\left(\partial_{i}\eta_{j}\right)^{\ast}+\kappa_{2}
\left(\partial_{i}\eta_{i}\right)
\left(\partial_{j}\eta_{j}\right)^{\ast}+
\kappa_{3}\left(\partial_{i}\eta_{j}\right)
\left(\partial_{j}\eta_{i}\right)^{\ast}+\kappa_{4}\left(\partial_{z}
\eta_{j}\right)
\left(\partial_{z}\eta_{j}\right)^{\ast}\}
\end{array}
\end{equation}

\noindent
with the order parameter given by the 2D vector

\begin{equation}
\label{e18}
\bar{\eta}=(\eta_{1},\eta_{2})
\end{equation}

\noindent
that is for 

\begin{equation}
\label{e19}
\psi\left({\bf k}\right)=\eta_{1}\phi_{1}\left({\bf k}\right)+
\eta_{2}\phi_{2}\left({\bf k}\right)
\end{equation}

\noindent
in the case of even parity (Eq. (\ref{e4})) and 

\begin{equation}
\label{e19a} 
d_{\mu}\left({\bf k}\right)=\eta_{1}\varphi_{1}^{\mu}\left({\bf k}\right)+
\eta_{2}\varphi_{2}^{\mu}\left({\bf k}\right)
\end{equation}

\noindent
for an odd parity (Eq. (\ref{e5})),   
where $ \phi_l\left({\bf k}\right)$ and $\varphi_{l}^{\mu}\left({\bf k}\right)$  
are the polynomials listed in Tabs. I-III. 
In the spin-triplet states we have assumed that the ${\bf d}$ vector  
is real and oriented along the crystal direction $\hat{c}$ 
($\hat{z}$-axis) by strong spin-orbit coupling in accordance with 
Sauls' scenario. \cite{1} 
This conjecture leads to the identical forms of the GL functionals for
both even  and odd states. \cite{24} 

Expressions for the GL coefficients are straightforward to obtain
through expansion of the Luttinger-Ward functional for the BCS
superconductor in powers of the order parameter, and identification
of the prefactors of the various invariants present in the phenomenological
form in Eq. (18).  One finds for the case of even parity

\begin{equation}
\label{e20}
\begin{array}{l}
\D\alpha=N_0\ln\frac{T}{T_c}\\
\\
\beta_{1}=2\beta_{2}=2\beta_{BCS}\left<\phi^2_i\left({\bf k}\right)
\phi^2_j\left({\bf k}\right)\right>\\
\\
\gamma_{1}=\frac{2}{3}\gamma_{2}=2\gamma_{BCS}
\left<\phi^2_i\left({\bf k}\right)\phi^4_j\left({\bf k}\right)\right>\\
\\
\kappa_{1}=\kappa_{wc}\left<v_i^2 \phi^2_j\left({\bf k}\right)\right>\\
\\
\kappa_{2}=\kappa_{3}=\kappa_{wc}
\left<v_i v_j\phi_i\left({\bf k}\right)\phi_j\left({\bf k}\right)\right>\\
\\
\kappa_{4}=\kappa_{wc}
\left<v_z^2\phi_i^2\left({\bf k}\right)\right>\\
\\
\kappa_{123}=\kappa_{wc}\left<v_i^2\phi_i^2
\left({\bf k}\right)\right>
\end{array}
\end{equation}

\noindent
where it is to be understood that \mbox{$i\neq j$} \mbox{$(i,j=x,y)$}, 
$\kappa_{123}=\kappa_1+\kappa_2+\kappa_3$, 
and the constants are given by  

\begin{equation}
\label{e21}
\begin{array}{l}
\D\beta_{BCS}=\frac{7}{16}\zeta\left(3\right)
\frac{N_0}{\left(\pi T_{c}\right)^2}\\
\\
\D\gamma_{BCS}= -\frac{31}{128}\zeta\left(5\right) 
\frac{N_0}{\left(\pi T_{c}\right)^4}\\
\\
\D\kappa_{wc}=\frac{7}{16}\zeta\left(3\right)
\frac{N_0}{\left(\pi T_c\right)^2} 
\end{array}
\end{equation}

\noindent
For odd parity these expression are
analogous where we have to take into account that a product $ \phi_i \phi_j $ 
is replaced by a scalar product $ \bar{\varphi}_i \cdot \bar{\varphi}_j $.

Knowledge of the Fermi surface and of the basis functions for the various
representations now allows one to evaluate, in principle, all coefficients
of interest (Eq. (\ref{e20})).  There are several hidden problems in this analysis.  
The most serious has been alluded to above, namely our lack of knowledge
of the relative pairing weight on each of the sheets of the Fermi surface.
As before, we approach this difficulty by assuming that pairing
takes place either on the $\Gamma_3$ or the $\Gamma_2$ sheets of the
Fermi surface, and discuss both cases separately.  The second has 
to do with the choice of basis functions.  We compare two different methods 
and calculate the coefficients for the expansion of the order parameter 
in the spherical harmonics and in the FS harmonics. Another complication 
arises from an infinite number of functions belonging to each irreducible   
representation in a crystal.  
Thus our choice of the $ \phi_l(\bf k)$ and $ \bar{\varphi}_l
({\bf k}) $ is perforce somewhat arbitrary.
In this work we generally work with  the lowest order polynomial 
functions for each representation.  We can test to see
whether the choice of higher order polynomials will greatly alter 
the size of the coefficients in Eq. (\ref{e20}) (it does not), but a more 
quantitative approach
would allow the order parameter $\hat\Delta_k$ to adjust 
to the given angle-resolved density of states by taking a general
linear combination of many such functions.  This would have, for the FS harmonic 
representation particularly, the 
virtue of allowing for smoothing of singularities which inevitably
result from the rapid changes of ${\bf v_k}$ over a complicated
Fermi surface.  Working with a single function 
$ \phi_l ({\bf v}_{{\bf k}}) $ and $ \varphi^{\mu}_l ({\bf v}_{{\bf k}}) $  
is sufficient for our purposes, however, since the coefficients in Eq. (\ref{e20}) 
are always integrated over the full FS under consideration, and only
the degree of hexagonal anisotropy should be important.

A special role in the topology of the $H-T$ phase diagram of the 
the 2D representations is played by the coefficient $\kappa_2$ 
(Eqs. (\ref{e17}), (\ref{e20})) which  mixes the gradients of
the two components of the order parameter.  The solution of the 
eigenvalue problem determining the critical field lines in the 
$H-T$ plane is analogous to a Sch\"odinger equation for a charge
moving in a magnetic field.  The existence
of a tetracritical point has been shown therefore to correspond to
a level crossing, which cannot exist in the absence of a
conserved quantum number. 
The $\kappa_2$ mixing terms correspond to a level repulsion 
(or hybridisation) term in this analogy, and thus prevent a crossing
of the critical field lines for $\bf H \parallel \hat c$.

By examining Table III, it is easy to see from Eq. (22) that 
the coefficient $\kappa_2$ in a cylindrically symmetric system,  
where $\bf v \parallel \bf k$, vanishes for $E_{2u}$ symmetry whereas for 
the $E_{1}$ representations,  
$\kappa_2/\kappa_{wc}v_{F\perp}^2$ is of $ {\cal O} (1)$.   Similarly
it is easy to check that $\kappa_1/\kappa_{wc}v_{F\perp}^2$ is 
of $ {\cal O} (1)$ for both representations.  Sauls' argument \cite{1} 
is therefore that if hexagonal anisotropy effects are weak $(\kappa_2\ll
\kappa_1)$, the critical fields may come so close as to mimic
an apparent tetracritical point, even for $\bf H \parallel \hat c$.  To
test this proposition, it is interesting to know not simply whether 
$\kappa_2$ is in fact much smaller than $\kappa_1$ when evaluated
over the ${\rm UPt}_3$ Fermi surface, but whether or
not this result is "accidental" or due to the near cylindrical symmetry
of the Fermi surface.  Is the change in $\kappa_2$ due to hexagonal
anisotropy in fact small?  Sauls proposes that it is, and gives a 
perturbative analysis for a band structure of type given by Eq. (\ref{e16}), 
assuming $a_3$
small.  In this case one finds $\kappa_2/\kappa_1 \simeq 0.1$ .\cite{29}  
It is however not clear
that a perturbative analysis is applicable. Here we calculate 
$\kappa_1$, $\kappa_2$ numerically for the band given by Eq. (\ref{e16}), 
and display the results 
as a function of the anisotropy parameter $a_3$ in Fig. 4. 
In the calculation presented in Fig. 4a the order parameter was expressed  
in the spherical harmonics, whereas Fig. 4b shows the result for 
the FS harmonic representation of the gap function.   
The $\Gamma_3$ Fermi sheet is given by $a_3/(\varepsilon_F c^6)=0.016$ 
(Table. IV) and the corresponding $\kappa_1$, $\kappa_2$ values are 
marked with a dotted line in Fig. 4.  
We read from Fig. 4a that $\kappa_2/\kappa_1\simeq 0.46$ 
for the order parameter expressed by the spherical harmonics 
and from Fig. 4b that $\kappa_2/\kappa_1\simeq 0.73$ for the FS harmonic 
expansion of the order parameter. Because of the cylindrical symmetry of  
$\Gamma_2$ FS, the
 $\kappa_2$ coefficient calculated over this FS sheet vanishes. 

\section{\bf Conclusions}
     We have tried to take crude features of the $UPt_3$ Fermi surface into
account within a weak-coupling theory of unconventional superconductivity
to see which order parameter symmetries are favored by Fermi surface
structure alone.  In the current philosophy,
these are classified according to their angular momentum quantum numbers
in the  fictitious spherical system obtained when the
ellipsoidal and hexagonal deformations of the ``true" Fermi surface are
``turned off".  Within this scheme, we have examined states of 
$p$, and $d$, and $f$ symmetry.
Other important elements of the  physics, such as the actual nature
of the spin-orbit coupling and the range of the pairing, may also play major
roles, but have been neglected here.
The lack of knowledge of a microscopic pairing mechanism--and
consequent inability to specify the relative pairing weights on the
various Fermi surface sheets--prevents us
from drawing firm conclusions from an analysis of this type, but 
our results are  suggestive.  
\vskip .2cm
It is  worth observing that
the few representations 
we have found to be stabilized
by Fermi surface anisotropy are precisely those under active consideration
as candidates for the $UPt_3$ order parameter: the two-dimensional
representations $E_{1g}$ and $E_{2u}$ and the 1D representations
$A_{1u}$ and $A_{1g}$.
Both of the 1D representations found are of 
interest in the accidentally degenerate
representation scenario of Garg and co-workers,\cite{3} which requires, however,
mixing of  nearly degenerate
$A$-type and $B$-type representations.  All of the latter are 
disfavored by the Fermi surface deformations we consider; our results do
not therefore lend support to this model.
The trivial representation $A_{1g}$ corresponds to basis functions with the full
symmetry of the Fermi surface, which typically are fully gapped.  
Such functions  
will be strongly supressed by local Coulomb 
interactions, which we do not
consider here; they are by themselves ruled out by experiment.   
Exceptions occur if  the system condenses 
in an $A_{1g}$ state with nodes, or mixes with another representation 
supporting basis functions with nodes, as in the Zhitomirskii-
Luk'yanchuk scenario.\cite{5}  The near degeneracy found in Figure 2b) for the $A_{1g}$ and
$E_{1g}$ states for $t\simeq 0.85$, which is very close to the best
fit to the $\Gamma_3$ Fermi surface sheet ($t=1$), is some support for
this picture.  
\vskip .2cm
We have attempted to account for our ignorance of the true distribution of
the pair weights over the various Fermi surface sheets by examining
models with pairing on the two sheets with highest density of states
according to de Haas-van Alphen measurements, namely $\Gamma_2$ and
$\Gamma_3$.  As $\Gamma_2$ is nearly ellipsoidal but $\Gamma_3$ has
strong hexagonal deformations, this choice has the
additional virtue of crudely separating hexagonal and ellipsoidal variations.
This may be of significance for current theories: for example, the hexagonal
deformation appears to be important in stabilizing the $E_{2u}$ representation.
Confirmation of the $E_{2u}$ model would therefore be indirect evidence for
pairing primarily on the $\Gamma_3$ sheet.  Such conclusions could be used
as guides for microscopic pairing theories.
\vskip .2cm
The scenario based on the $E_{2u}$ representation due to Sauls\cite{1}
is based on the assumption that the level repulsion term in the Schr\"odinger-like
equation for the critical field lines vanishes due to the proximity of
the system to cylindrical symmetry.  This term is proportional to the
ratio of  GL gradient coefficients $(\kappa_2/\kappa_1)^2$,\cite{29} which vanishes
in this limit for the $E_{2u}$ state.  We have calculated these
coefficients over our model Fermi surfaces fit to LDA calculations
and dHvA measurements, and find that
this term is not small, but varies between .21 and .53, depending on our
assumptions regarding the exact form of the basis functions.  
While the parametrization
of the Fermi surface we have adopted is extremely simplistic, it has the
virtue that one can discuss the proximity to cylindrical symmetry.   What
Figure 4 suggests is that the $UPt_3$ system may not really be regarded
as close to cylindrically symmetric, and that therefore if calculations 
of the GL coefficients over the true Fermi 
surface produce a smaller value,\cite{29}  it should be regarded
as coincidence.   This undermines somewhat the attractiveness of the Sauls
$E_{2u}$ scenario, but of course does not constitute   proof against it.
One way in which this conclusion could be avoided is to have
contributions to $\kappa_2$ which change sign on different Fermi 
surface sheets, leading to a smaller effective $\kappa_2$.  This
effect is indeed found in a calculation with the full $UPt_3$ LDA
Fermi surface and a model with weight distributed equally
over all sheets.\cite{Normanprivate1}
\vskip .2cm
\section*{\bf Acknowledgments}
We are grateful to M. Norman and J. A. Sauls for many helpful discussions. 
One of us (G.H.) was partially supported by the Fulbright Foundation. 
Furthermore, M.S. would like to thank the Swiss Nationalfonds for
support by a PROFIL fellowship.

\newpage
\begin{table}
\begin{center}
\caption{Basis functions for the gap function:
p-wave}
\begin{tabular}{||c|ccc|ccccc|ccc||} 
\hline 
$l$ && $Irreducible$ &&&& $\bar{\varphi}_l({\bf k})$ &&&& $degenerate$ & \\
&& $representation\;\Gamma$ &&&&&&&& $with$ &\\
\hline
1&& $A_{1u}$ &&&& $\hat{{\bf z}} k_z$ &&&&& \\
2&& $A_{1u}$ &&&& $\hat{{\bf x}} k_x + \hat{{\bf y}} k_y$ &&&& 
$E_{1u}({\bf d}\parallel\hat{z})$ & \\
3&& $A_{2u}$ &&&& $\hat{{\bf x}} k_y - \hat{{\bf y}} k_x$ &&&& 
$E_{1u}({\bf d}\parallel\hat{z})$ & \\
4&& $E_{1u}$ &&&& $\hat{{\bf z}} k_x$ &&&&& \\
5&&&&&& $\hat{{\bf z}}k_y$ &&&&& \\
6&& $E_{1u}$ &&&& $\hat{{\bf x}} k_z $ &&&& $A_{1u}({\bf d}\parallel\hat{z})$ & \\
7&&&&&& $\hat{{\bf y}} k_z$ &&&&& \\
8&& $E_{2u}$ &&&& $\hat{{\bf x}}k_x-\hat{{\bf y}}k_y$ &&&& 
$E_{1u}({\bf d}\parallel\hat{z})$ &\\
9&&&&&& $\hat{{\bf x}} k_y+\hat{{\bf y}}k_x$ &&&&& \\
\end{tabular}
\end{center}
\end{table}

\begin{table}
\begin{center}
\caption{Basis functions for the gap function:
d-wave}
\begin{tabular}{||c|ccc|ccccc||} 
\hline 
$l$ && $Irreducible$ &&&& $\phi_l({\bf k})$ && \\
&& $representation\;\Gamma$ &&&&&& \\
\hline
1&& $A_{1g}$ &&&& $2k^2_z - k^2_x -k^2_y$ && \\
2&& $E_{1g}$ &&&& $k_x k_z$ && \\
3&&          &&&& $k_y k_z $ && \\
4&& $E_{2g}$ &&&& $k^2_x - k^2_y$ && \\
5&&          &&&& $2 k_x k_y$ && \\  
\end{tabular}
\end{center}
\end{table}

\begin{table}
\begin{center}
\caption{Basis functions for the gap function:
f-wave}
\begin{tabular}{||c|ccc|ccccc|ccc||} 
\hline 
$l$ && $Irreducible$ &&&& $\bar{\varphi}_l({\bf k})$ &&&& $degenerate$ & \\
&& $representation\;\Gamma$ &&&&&&&& $with$ &\\
\hline 
10&& $A_{1u}$ &&&& $\hat{{\bf z}} k_z (2k^2_z - 3 k^2_x - 3 k^2_y)$ &&&&& \\
11&& $A_{1u}$ &&&& $(\hat{{\bf x}}k_x + \hat{{\bf y}}k_y)
(4k^2_z - k^2_x - k^2_y)$ &&&& $E_{1u}({\bf d}\parallel\hat{z})$ &\\
12&& $A_{2u}$ &&&& $(\hat{{\bf x}}k_y-\hat{{\bf y}}k_x) 
(4k^2_z - k^2_x - k^2_y)$ &&&& $E_{1u}({\bf d}\parallel\hat{z})$ &\\
13&& $B_{1u}$ &&&& $\hat{{\bf z}} (k^3_x - 3 k^2_y k_x)$ &&&&& \\
14&& $B_{1u}$ &&&& $\hat{{\bf x}}k_z(k^2_x - k^2_y) +2 \hat{{\bf y}}
k_x k_y k_z$ &&&& $E_{2u}({\bf d}\parallel\hat{z})$ &\\
15&& $B_{2u}$ &&&& $\hat{{\bf z}} (k^3_y - 3 k^2_x k_y)$ &&&&& \\
16&& $B_{2u}$ &&&& $\hat{{\bf y}} k_z (k^2_x-k^2_y) + 2\hat{{\bf x}}
k_x k_y k_z$ &&&& $E_{2u}({\bf d}\parallel\hat{z})$ &\\
17&& $E_{1u}$ &&&& $\hat{{\bf z}} k_x (4k^2_z - k^2_x - k^2_y)$ &&&&& \\
18&&          &&&& $\hat{{\bf z}} k_y (4k^2_z - k^2_x - k^2_y)$ &&&&& \\
19&& $E_{1u}$ &&&& $\hat{{\bf x}} k_z (2k^2_z - 3 k^2_x - 3 k^2_y)$ &&&& 
$A_{1u}({\bf d}\parallel\hat{z})$ &\\
20&&          &&&& $\hat{{\bf y}} k_z (2k^2_z - 3 k^2_x - 3 k^2_y)$ &&&&& \\
21&& $E_{1u}$ &&&& $\hat{{\bf x}} k_z (k^2_x -k^2_y) - 2\hat{{\bf y}}
k_x k_y k_z$ &&&& $E_{2u}({\bf d}\parallel\hat{z})$ &\\
22&&          &&&& $\hat{{\bf y}} k_z(k^2_x -k^2_y) - 2\hat{{\bf x}}
k_x k_y k_z$ &&&&& \\
23&& $E_{2u}$ &&&& $\hat{{\bf z}} k_z (k^2_x - k^2_y)$ &&&&& \\
24&&          &&&& $2 \hat{{\bf z}} k_x k_y k_z$ &&&&& \\
25&& $E_{2u}$ &&&& $(\hat{{\bf x}} k_x - \hat{{\bf y}} k_y)
(4k^2_z - k^2_x - k^2_y)$ &&&& $E_{1u}({\bf d}\parallel\hat{z})$ &\\
26&&          &&&& $(\hat{{\bf x}} k_y + \hat{{\bf y}} k_x)
(4k^2_z - k^2_x - k^2_y)$ &&&&& \\
27&& $E_{2u}$ &&&& $\hat{{\bf x}} (k^3_x - 3 k^2_y k_x)$ &&&& 
$B_{1u}({\bf d}\parallel\hat{z})$ & \\
28&&          &&&& $\hat{{\bf y}} (k^3_x - 3 k^2_y k_x)$ &&&&& \\
29&& $E_{2u}$ &&&& $\hat{{\bf x}} (k^3_y - 3 k^2_x k_y)$ &&&& 
$B_{2u}({\bf d}\parallel\hat{z})$ &\\
30&&          &&&& $\hat{{\bf y}} (k^3_y - 3 k^2_x k_y)$ &&&&& \\
\end{tabular}
\end{center}
\end{table}

\newpage
\begin{table}
\begin{center}
\caption{The functions and the coefficients used in fitting $\Gamma_{2}$
and $\Gamma_{3}$ energy bands, $c=4.9027\AA$ is the z-axis lattice 
constant (Ref. 35).}
\begin{tabular}{||c|ccc|ccc|ccc|ccc||}
\hline 
$i$ && $a_i$ &&& $\Gamma_{2}$ &&& $\Gamma_{3}$ &&& $f_i({\bf k})$ &\\ \hline
1&& $a_1/(\varepsilon_F c^2)$ &&& $0.466$ &&& $0.157$ &&& $k_{z}^{2}$ &\\
2&& $a_2/(\varepsilon_F c^2)$ &&& $0.951$ &&& $0.241$ &&& $k_{x}^{2}+k_{y}^{2}$ &\\
3&& $a_3/(\varepsilon_F c^6)$ &&& $0$ &&& $0.016$ &&& 
$\left(k_{y}^{3}-3k_{x}^{2}k_{y}\right)^{2}$ &\\  
\end{tabular}
\end{center}
\end{table}

\newpage
\section*{Figure Captions}

\noindent
Fig. 1. $\Gamma_3$ (solid line) and $\Gamma_2$ (dashed line) Fermi sheet plot 
in the symmetry plane: a) $K\Gamma M$, b) $L\Gamma M$, c) $H\Gamma K$.\\

\noindent
Fig. 2. The hexagonal crystal field $t$ effect on the critical temperature of 
the a) p-wave states, b) d-wave states, c) f-wave states.  $t=0$ corresponds
to isotropic, $t=1$ to best fit dHvA-LDA $\Gamma_3$ Fermi surface.\\

\noindent
Fig. 3. The cylindrical  crystal field $t$ effect on the critical temperature
of the a) p-wave states, b) d-wave states, c) f-wave states.
 $t=0$ corresponds
 to isotropic, $t=1$ to best fit dHvA-LDA $\Gamma_2$ Fermi surface.\\ 

\noindent
Fig. 4. Normalized $\kappa_1$ (solid line) and $\kappa_2$ (dashed line)  
stiffness coefficients as the functions of the normalized FS hexagonal 
anisotropy parameter $a_3$: a) spherical harmonics representation, 
b) FS harmonics representation.  The dotted line corresponds to the
value $a_3/\epsilon_Fc^6=0.016$ obtained by a fit to the $\Gamma_3$ Fermi surface
sheet.\\

\newpage
\begin{center}
\begin{figure}[p]
\parbox{15cm}{\epsfig{file=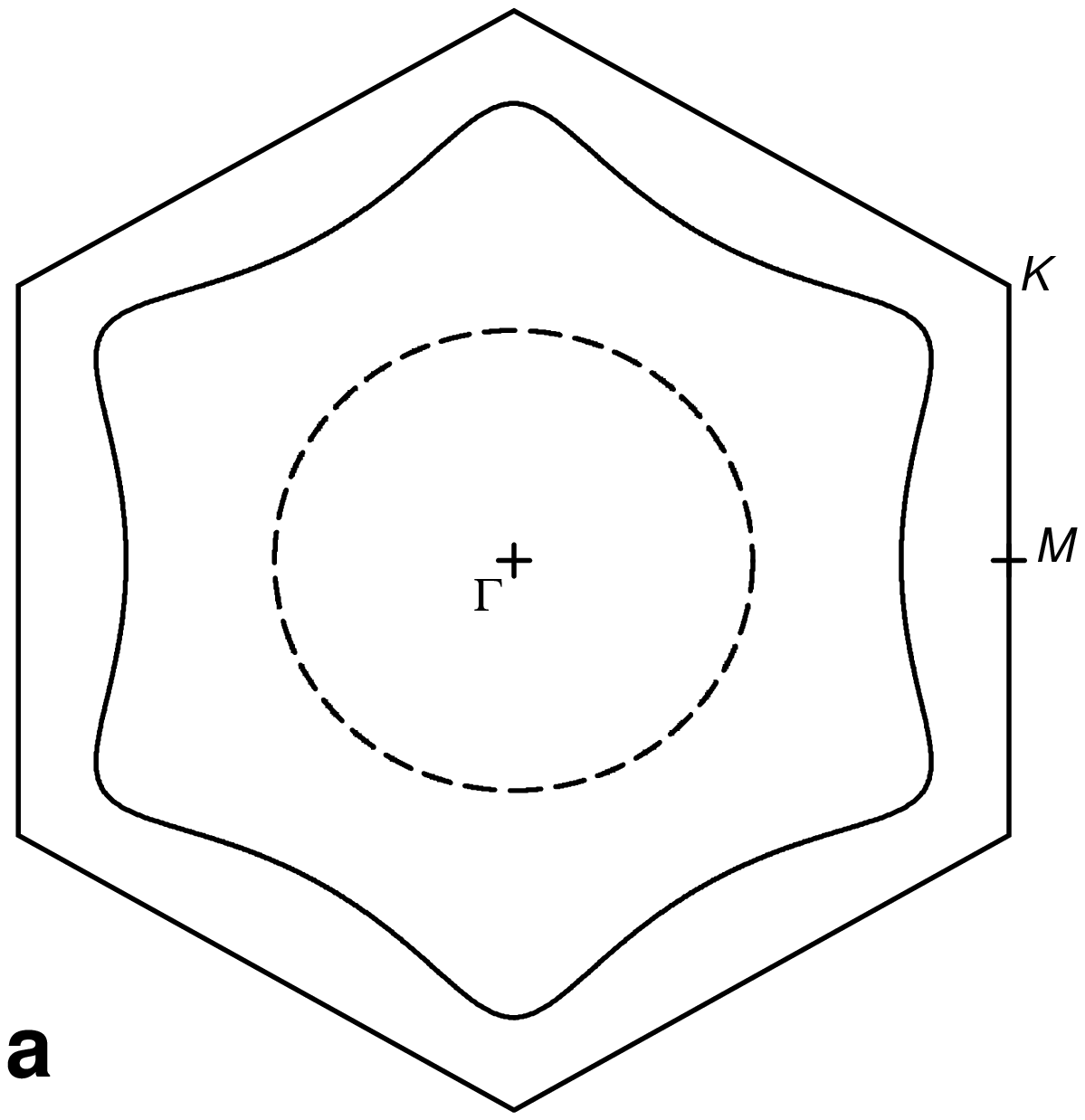,height=12.990381cm,width=15cm} }
\parbox{0.5cm}{\hfill} 
\end{figure}
\end{center}

\newpage
\begin{center}
\begin{figure}[p]
\parbox{15cm}{\epsfig{file=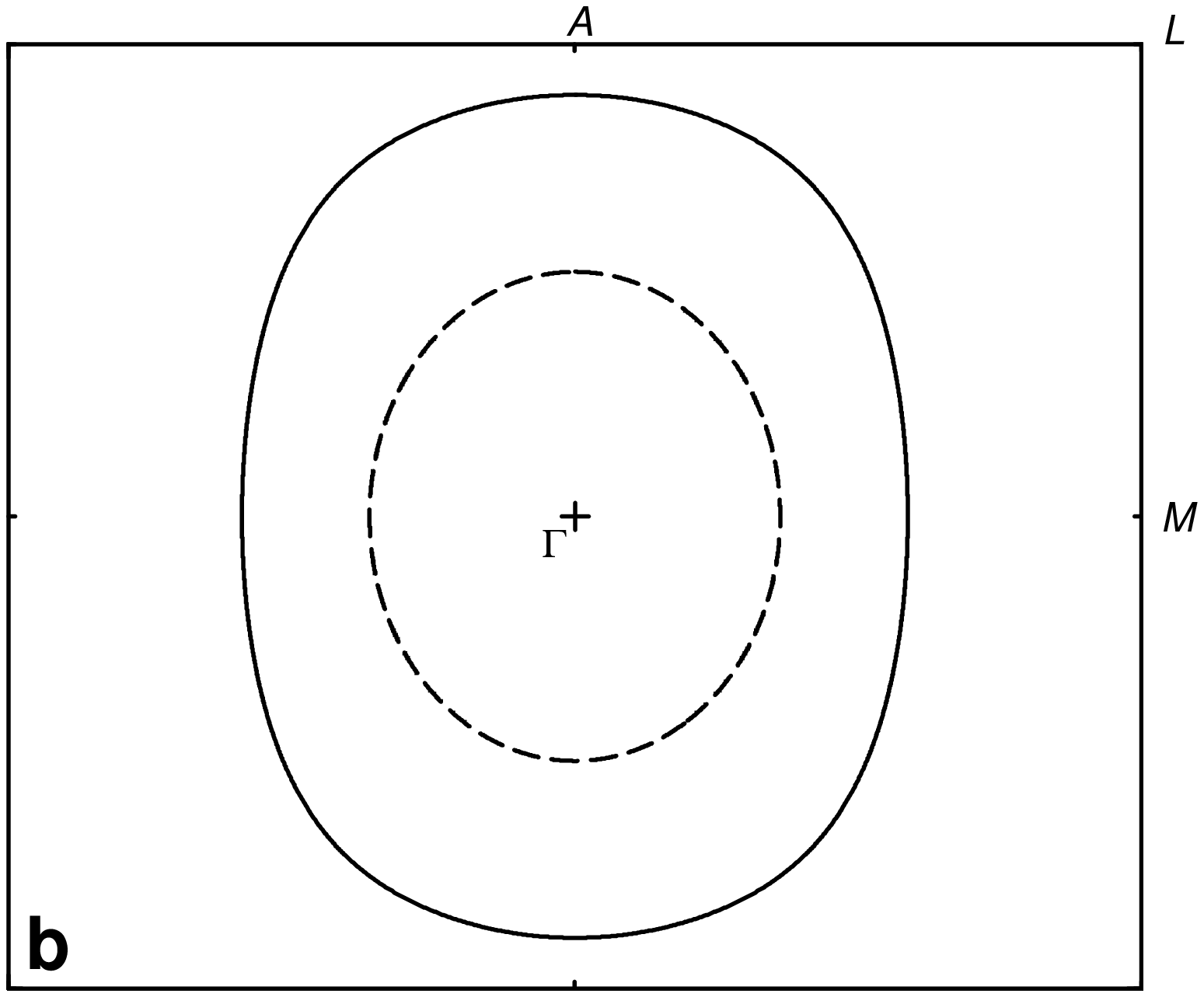,height=15cm,width=15cm} }
\parbox{0.5cm}{\hfill} 
\end{figure}
\end{center}

\newpage
\begin{center}
\begin{figure}[p]
\parbox{15cm}{\epsfig{file=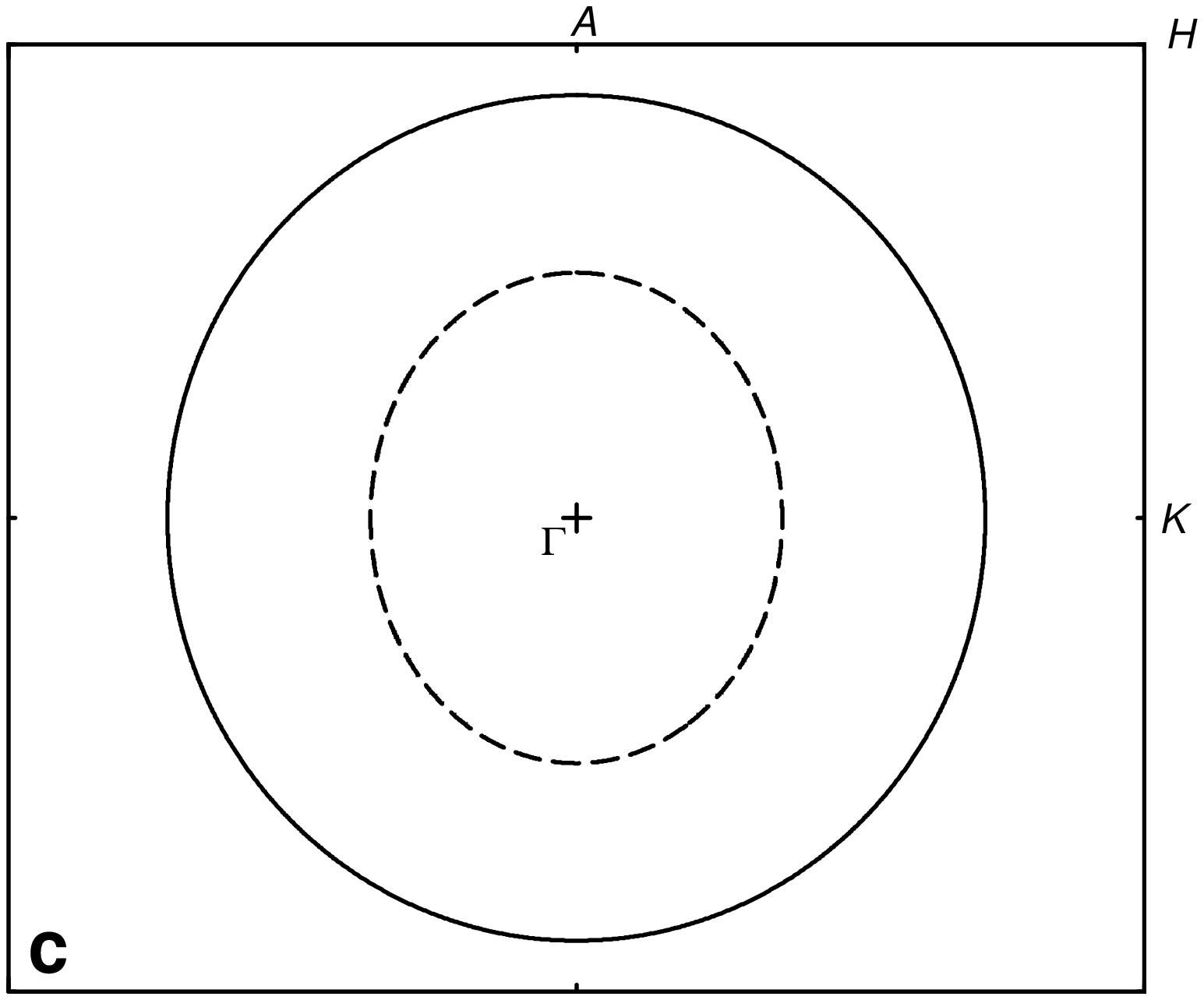,height=15cm,width=15cm} }
\parbox{0.5cm}{\hfill} 
\end{figure}
\end{center}

\newpage
\begin{center}
\begin{figure}[p]
\parbox{0.1cm}{\LARGE\vfill $$\nu'$$\vspace{3ex}\vfill }
\parbox{15cm}{\epsfig{file=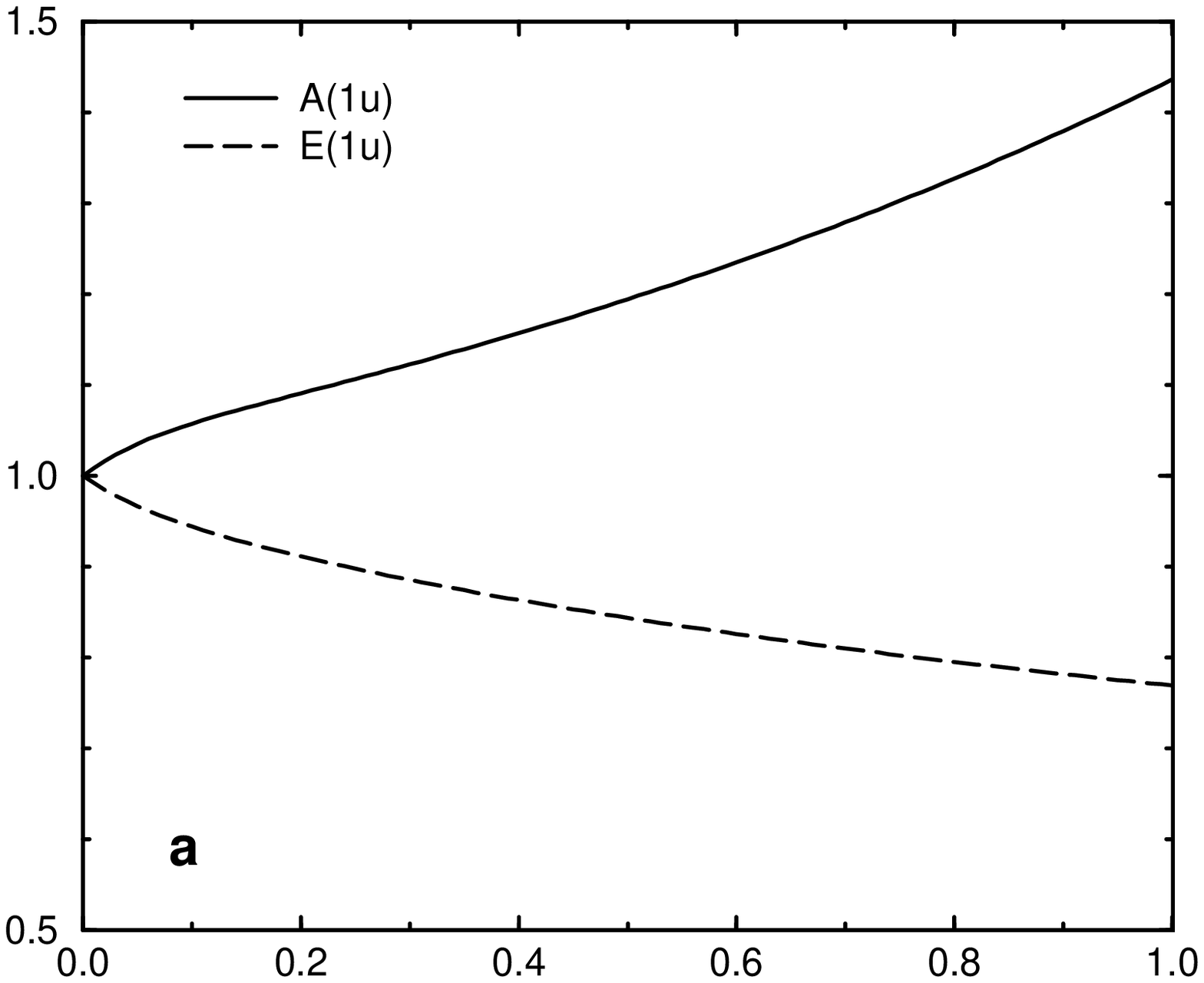,height=15cm,width=15cm} }
\parbox{0.5cm}{\hfill}
\parbox{18cm}{\LARGE\vspace{-6ex}\hfill $$t \;\;\;\;\;$$\hfill}
\end{figure}
\end{center}

\newpage
\begin{center}
\begin{figure}[p]
\parbox{0.1cm}{\LARGE\vfill $$\nu'$$\vspace{3ex}\vfill }
\parbox{15cm}{\epsfig{file=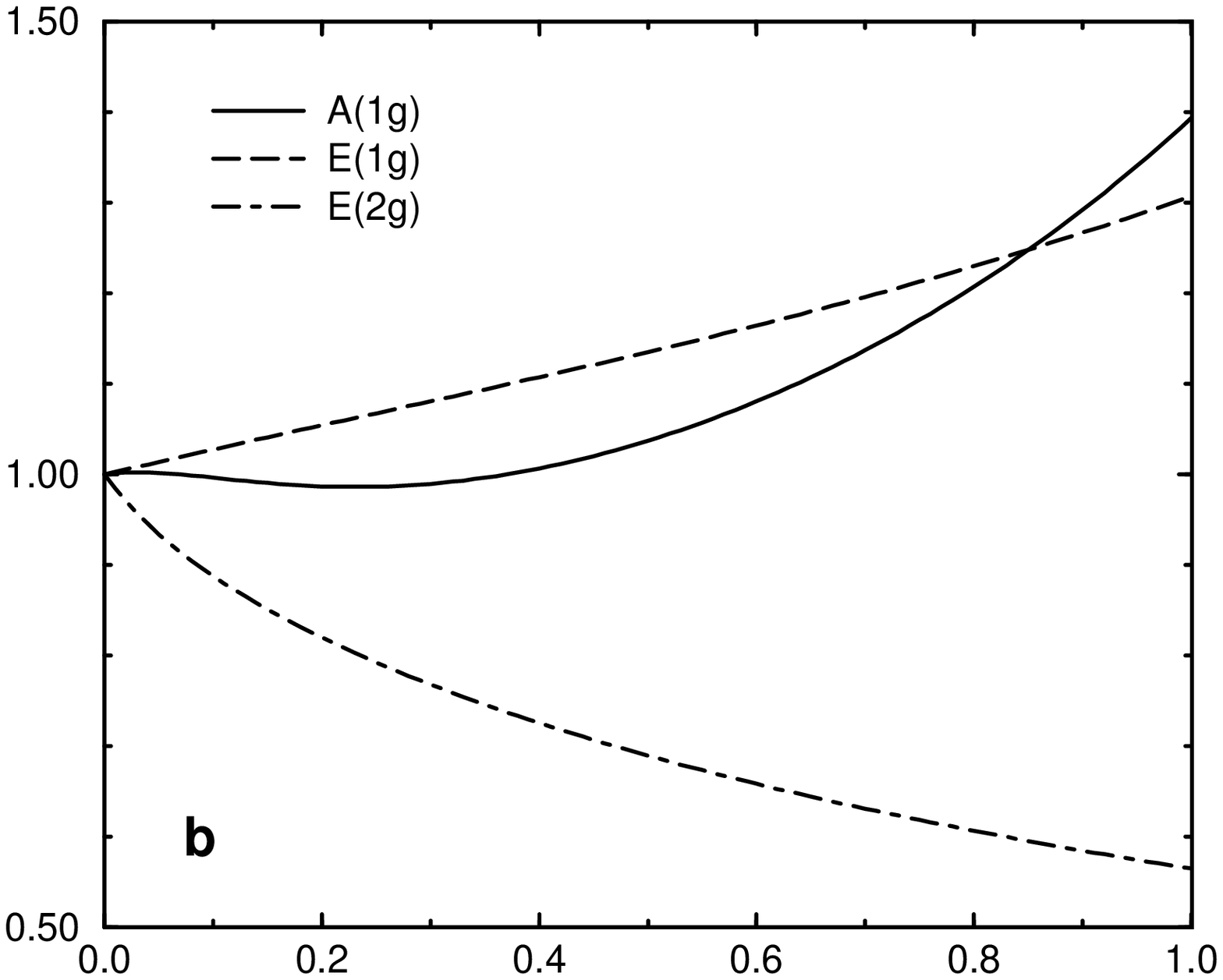,height=15cm,width=15cm} }
\parbox{0.5cm}{\hfill}
\parbox{18cm}{\LARGE\vspace{-6ex}\hfill $$t \;\;\;\;\;$$\hfill}
\end{figure}
\end{center}

\newpage
\begin{center}
\begin{figure}[p]
\parbox{0.1cm}{\LARGE\vfill $$\nu'$$\vspace{3ex}\vfill }
\parbox{15cm}{\epsfig{file=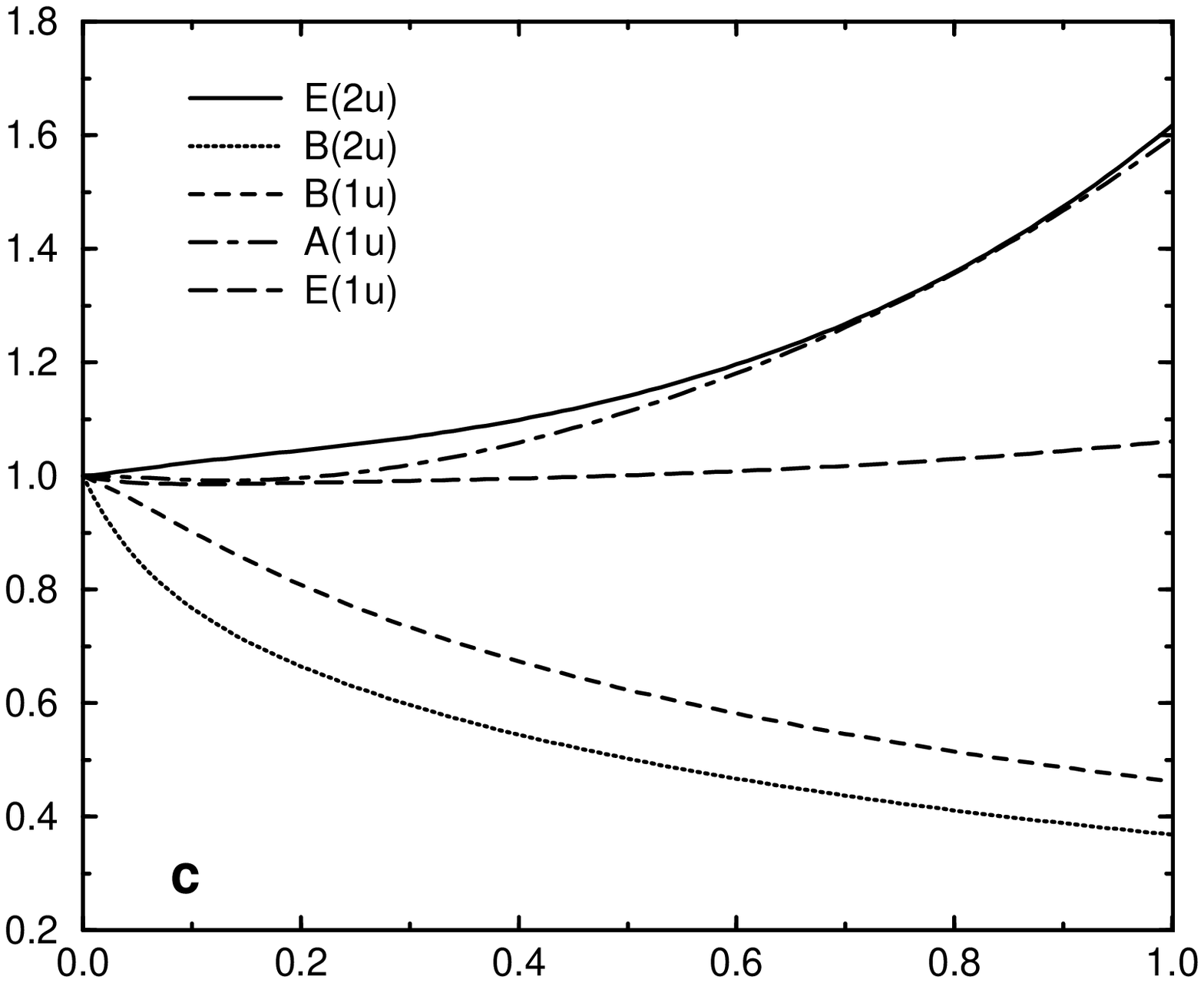,height=15cm,width=15cm} }
\parbox{0.5cm}{\hfill}
\parbox{18cm}{\LARGE\vspace{-6ex}\hfill $$t \;\;\;\;\;$$\hfill}
\end{figure}
\end{center}

\newpage
\begin{center}
\begin{figure}[p]
\parbox{0.1cm}{\LARGE\vfill $$\nu'$$\vspace{3ex}\vfill }
\parbox{15cm}{\epsfig{file=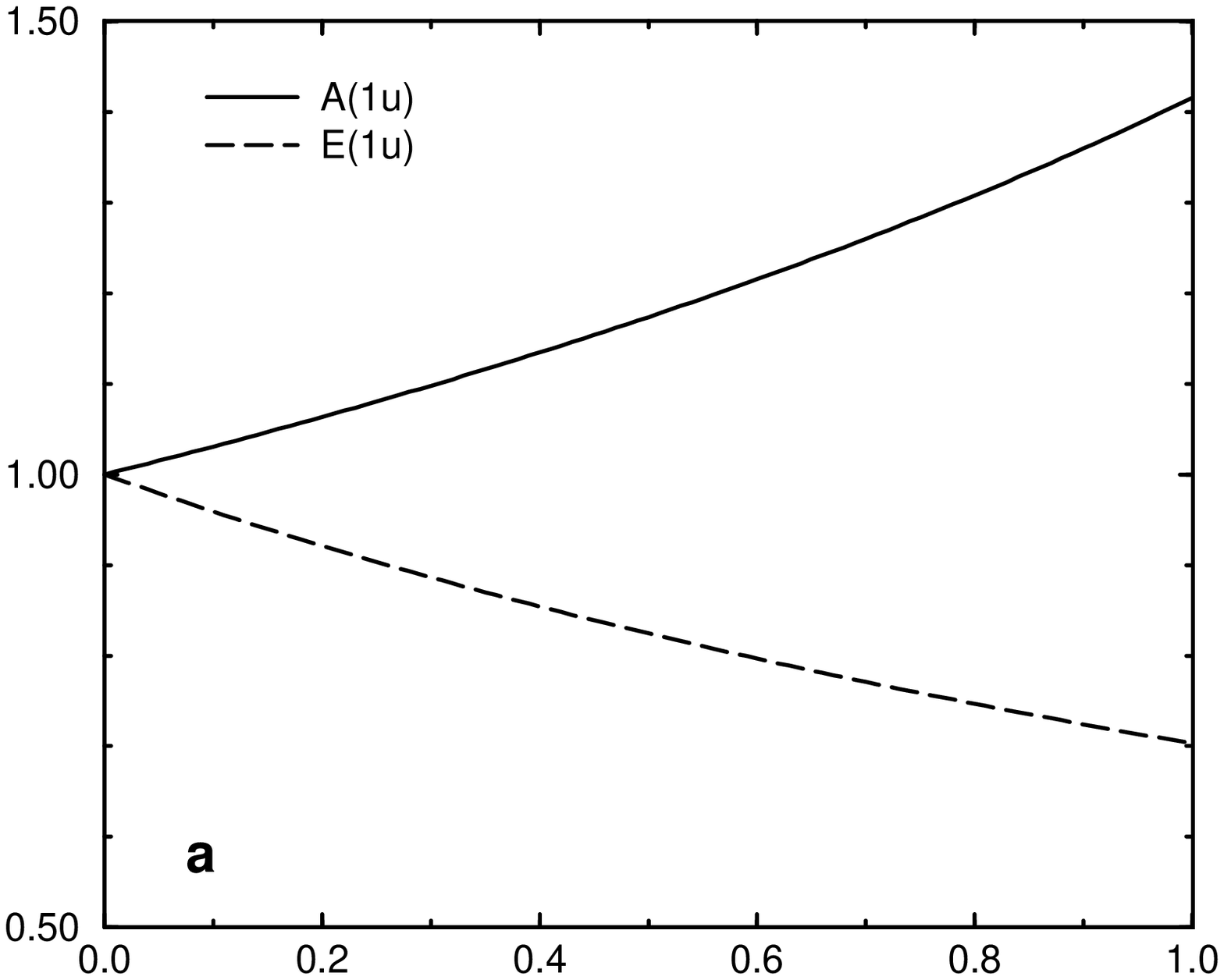,height=15cm,width=15cm} }
\parbox{0.5cm}{\hfill}
\parbox{18cm}{\LARGE\vspace{-6ex}\hfill $$t \;\;\;\;\;$$\hfill}
\end{figure}
\end{center}

\newpage
\begin{center}
\begin{figure}[p]
\parbox{0.1cm}{\LARGE\vfill $$\nu'$$\vspace{3ex}\vfill }
\parbox{15cm}{\epsfig{file=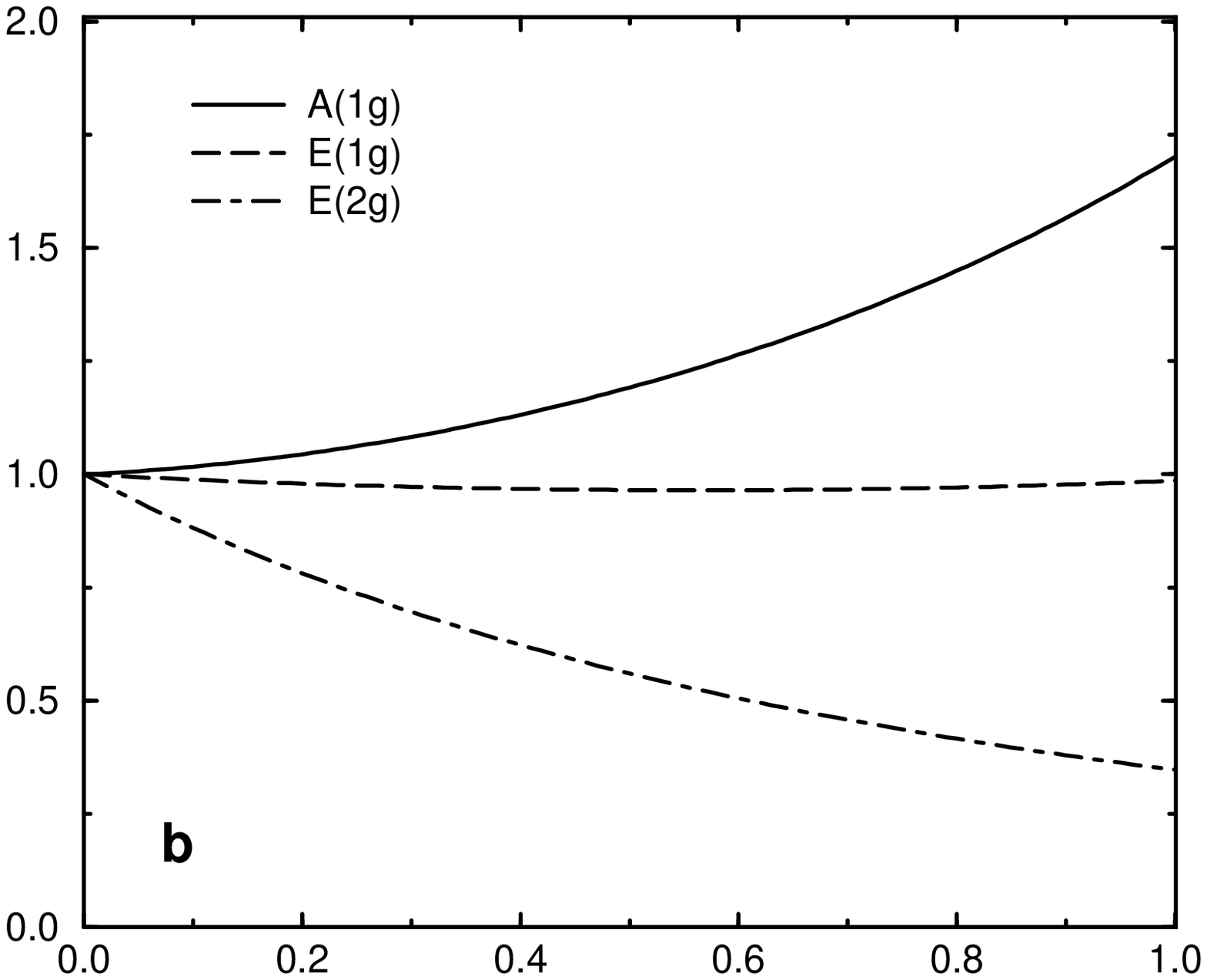,height=15cm,width=15cm} }
\parbox{0.5cm}{\hfill}
\parbox{18cm}{\LARGE\vspace{-6ex}\hfill $$t \;\;\;\;\;$$\hfill}
\end{figure}
\end{center}

\newpage
\begin{center}
\begin{figure}[p]
\parbox{0.1cm}{\LARGE\vfill $$\nu'$$\vspace{3ex}\vfill }
\parbox{15cm}{\epsfig{file=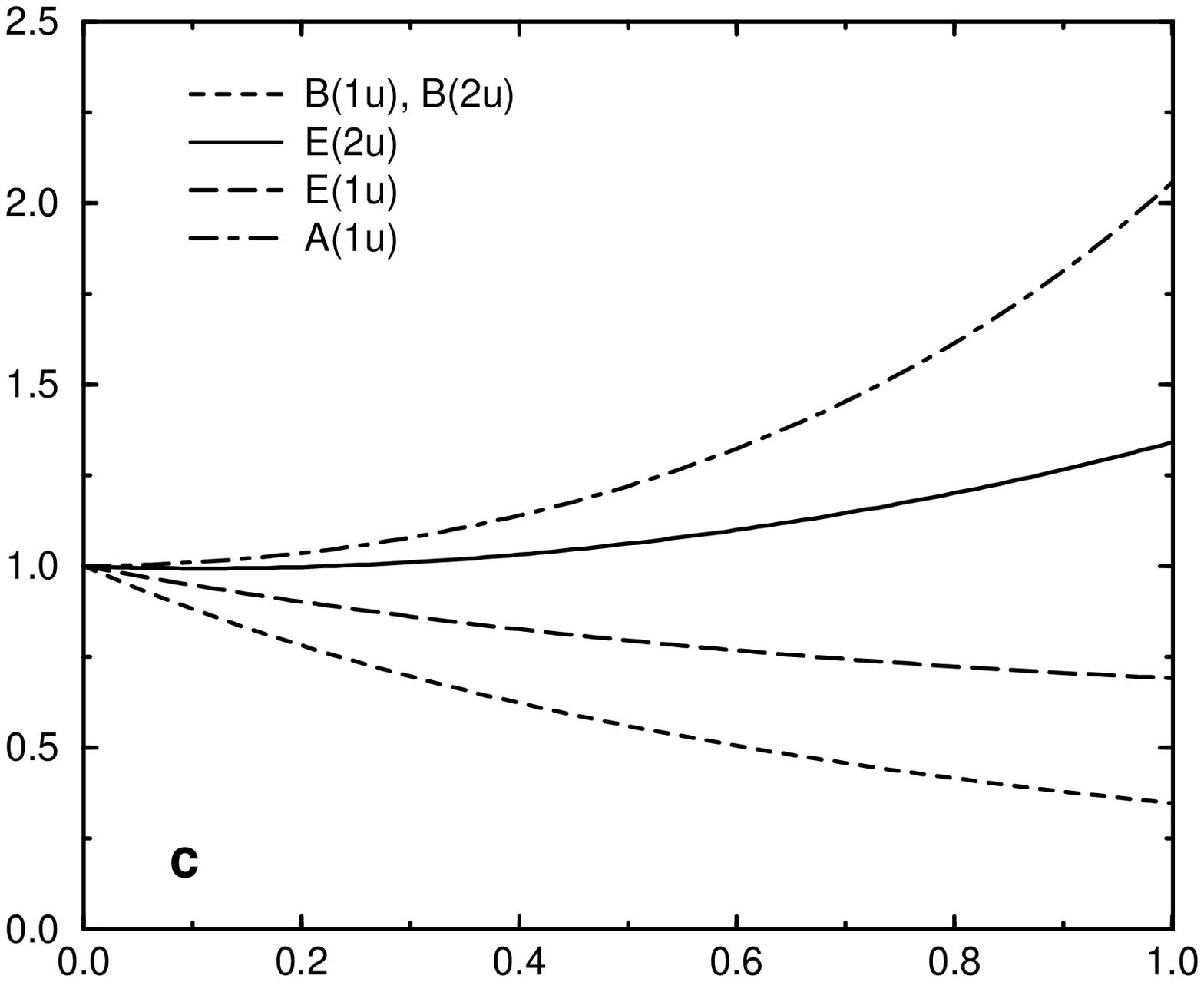,height=15cm,width=15cm} }
\parbox{0.5cm}{\hfill}
\parbox{18cm}{\LARGE\vspace{-6ex}\hfill $$t \;\;\;\;\;$$\hfill}
\end{figure}
\end{center}

\newpage
\begin{center}
\begin{figure}[p]
\parbox{1cm}{\LARGE\vfill $$\kappa_i/\kappa_{123}\;\;$$\vspace{2ex}\vfill }
\parbox{15cm}{\epsfig{file=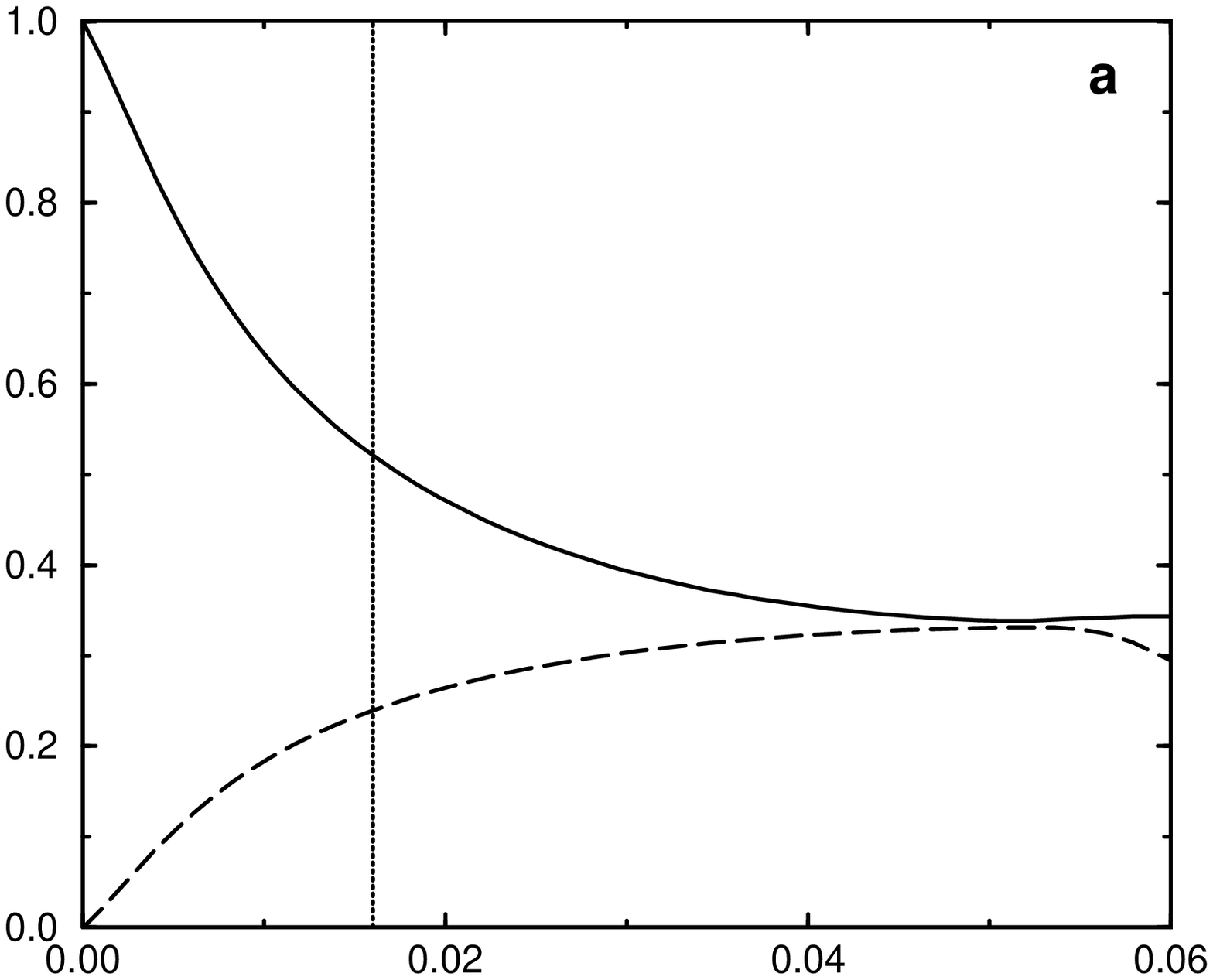,height=15cm,width=15cm} }
\parbox{0.5cm}{\hfill}
\parbox{18cm}{\LARGE\vspace{-6ex}\hfill $$\;\;\: a_3/(\varepsilon_F c^6)$$\hfill} 
\end{figure}
\end{center}

\newpage
\begin{center}
\begin{figure}[p]
\parbox{1cm}{\LARGE\vfill $$\kappa_i/\kappa_{123}\;\;$$\vspace{2ex}\vfill }
\parbox{15cm}{\epsfig{file=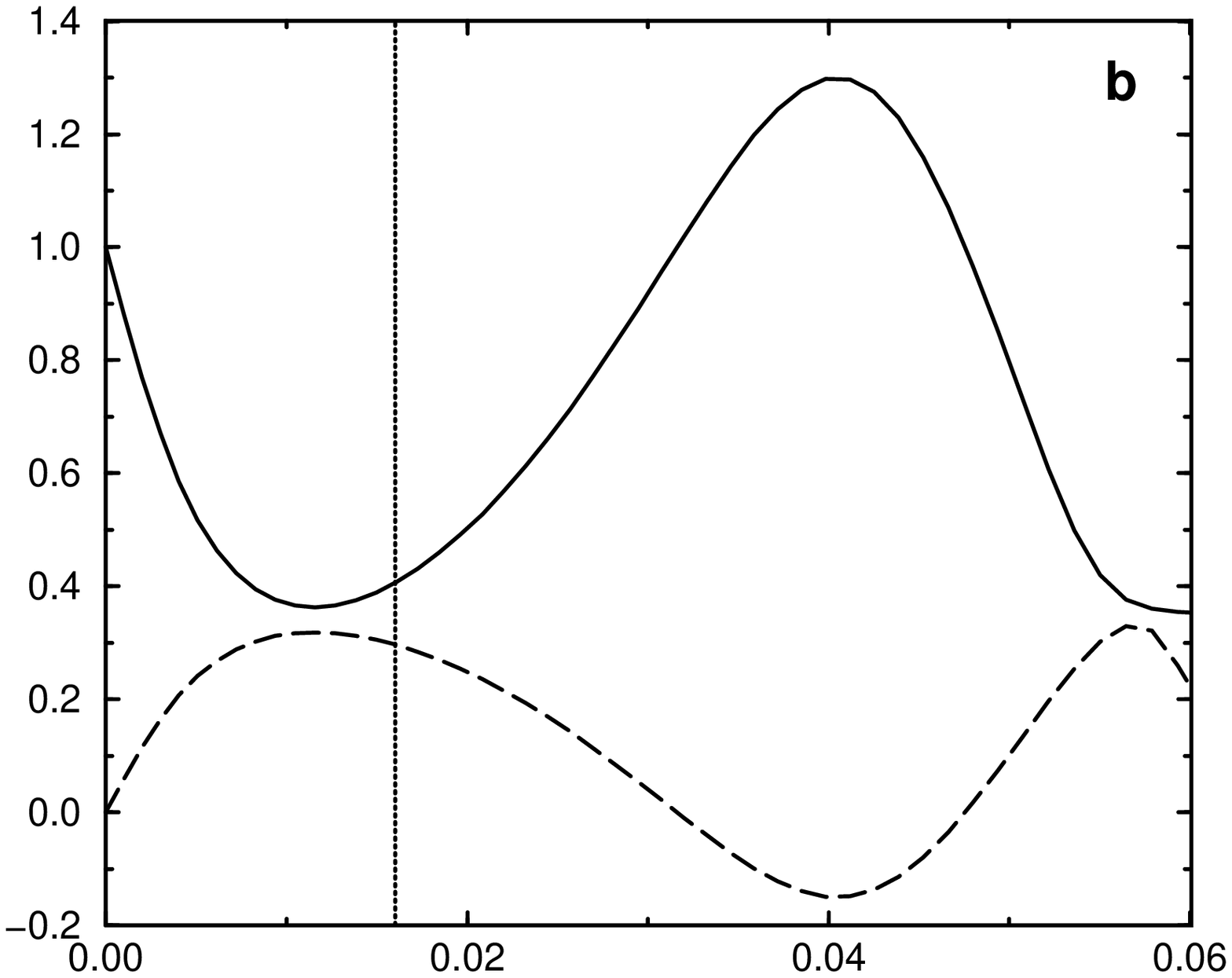,height=15cm,width=15cm} }
\parbox{0.5cm}{\hfill}
\parbox{18cm}{\LARGE\vspace{-6ex}\hfill $$\;\;\: a_3/(\varepsilon_F c^6)$$\hfill}
\end{figure}
\end{center}

\end{document}